# Thermal Stability Study of Transition Metal Perovskite Sulfides

Shanyuan Niu[†], JoAnna Milam-Guerrero[§], Yucheng Zhou[†], Kevin Ye[†], Boyang Zhao[†], Brent C. Melot[§], Jayakanth Ravichandran*[†][‡]

[†]Mork Family Department of Chemical Engineering and Materials Science, University of Southern California, Los Angeles, CA 90089, USA

[§]Department of Chemistry, University of Southern California, Los Angeles, CA 90089, USA

[‡]Ming Hsieh Department of Electrical Engineering, University of Southern California, Los Angeles, CA 90089, USA

*jayakanr@usc.edu

**Abstract**: Transition metal perovskite chalcogenides, a class of materials with rich tunability in functionalities, are gaining increased attention as candidate materials for renewable energy applications. Perovskite oxides are considered excellent *n*-type thermoelectric materials. Compared to oxide counterparts, we expect the chalcogenides to possess more favorable thermoelectric properties such as lower lattice thermal conductivity and smaller band gap, making them promising material candidates for high temperature thermoelectrics. Thus, it is necessary to study the thermal properties of these materials in detail, especially thermal stability, to evaluate their potential. In this work, we report the synthesis and thermal stability study of five compounds, $\alpha$-SrZrS$_3$, $\beta$-SrZrS$_3$, BaZrS$_3$, Ba$_2$ZrS$_4$, and Ba$_3$Zr$_2$S$_7$. These materials cover several structural types including distorted perovskite, needle-like, and Ruddlesden-Popper phases. Differential scanning calorimeter and thermo-gravimetric analysis measurements were performed up to 1200°C in air. Structural and chemical characterizations such as X-ray diffraction, Raman spectroscopy, and energy dispersive analytical X-ray spectroscopy were performed on all the samples before and after the heat treatment to understand the oxidation process. Our studies show that perovskite chalcogenides possess excellent thermal stability in air at least up to 600°C.

# I. Introduction

Rational design of new materials or identification of new functionalities in underexplored materials, especially semiconductors, has been a key contributor to electronic, photonic, and energy technologies. Transition metal perovskite chalcogenides (TMPCs), a class of materials with rich tunability and functionality, are currently of high interest for applications as solar cells and infrared detectors among other applications. TMPCs have a general chemical formula of $ABX_3$, where A is an alkaline earth metal such as Ba, Sr, Ca, B is a transition metal such as Ti, Zr, Hf, and X is a chalcogen such as S, Se. Specifically, the materials with early transition metals such as Ti, Zr, Hf on the B-site have recently been explored for optoelectronic and photonic applications.[1-10] Further exploration of the chemical and thermal stability for these materials, with respect to heating or exposure to moisture and air, is one of the key remaining questions to extent their potential for large-scale applications.[11]

TMPCs share a lot of exciting features with the well-studied oxide counterparts, including rich, tunable chemistry, high stability, and environmental friendly and earth abundant composition. Similar to the perovskite oxides, the valence band and the conduction band of TMPC are primarily composed of chalcogen *p* orbitals and transition metal *d* orbitals, respectively. High density of states (DOS) is thus expected from the combination of highly symmetric structure and degenerate transition metal *d* orbitals. These features are key functionalities for high temperature thermoelectric candidates.[12] In fact, perovskite oxides have been extensively studied for thermoelectrics.[13-18] However, the high lattice thermal conductivity and large band gaps limit their thermoelectric performance. With the replacement of oxygen with larger, heavier, and less electronegative chalcogen elements, TMPCs are expected to possess lower thermal conductivity and lower band gaps spanning IR to visible spectrum, thereby mitigating those issues in the oxide counterparts. Thus, it is also important to study the thermal properties of TMPCs to evaluate their potential for thermoelectrics.

In our previous studies, we have demonstrated high quality synthesis of polycrystalline TMPCs through solid state reaction in sealed ampoules,[9] as well as single crystal growth with chemical vapor transport and salt flux methods.[19-21] We have also evaluated their physical properties for photovoltaics and anisotropic Infrared photonic materials.[9,19-21] In this work, we will study the thermal stability of several transition metal perovskite related sulfides in air, including $\alpha$-SrZrS$_3$ ($\alpha$-SZS), $\beta$-SrZrS$_3$ ($\beta$-SZS), BaZrS$_3$ (BZS), Ba$_2$ZrS$_4$ (BZS214), and Ba$_3$Zr$_2$S$_7$ (BZS327).

## II. Materials and Methods

Materials Synthesis: Synthesis of all the samples were performed *via* solid state reaction in sealed quartz ampoules. Binary sulfides, elemental metal powders and sulfur pieces were used as precursors. As described in our previous work, the addition of small amount of iodine can effectively reduce the reaction time and avoid the laborious repeated cycles of grinding, pelletizing and heating. To avoid potential oxidation, the synthetic procedure was performed in a controlled atmosphere. The starting materials, barium sulfide powder (Sigma-Aldrich, 99.9%), strontium sulfide powder (Alfa Aesar, 99.9%), lanthanum sulfide powder (Alfa Aesar 99%), zirconium powder (STREM, 99.5%), sulfur pieces (Alfa Aesar 99.999%), and iodine pieces

(Alfa Aesar 99.99%) were stored and handled in an argon-filled glove box. Stoichiometric quantities of precursor powders with a total weight of ~0.5 g were ground and loaded into a quartz tube along with approximately 0.5-2 mg/cm$^3$ iodine inside the glove box. The quartz tube has an outside diameter of 19 mm and an inner diameter of 15 mm. The tube was capped with ultra-torr fittings and a bonnet needle valve to avoid exposure to the air during transfer. The tube was then evacuated with an anti-corrosive rotary vane pump to below 5 mTorr, and then sealed using a blowtorch with oxygen and natural gas combustion mixture without exposing the contents of the ampoule to air. All tubes were first dwelled at 400 °C to allow consumption of sulfur vapor and then ramped to the growth temperature with a ramp rate of 100 °C/h. All the samples were quenched to room temperature after the dwell time using a sliding furnace setup with a cooling rate around 100 °C/min. BZS, $\alpha$-SZS, $\beta$-SZS, BZS214, and BZS327 samples were held at 600 °C, 850 °C, 1100 °C, 1050 °C, and 1100 °C, respectively, for 60-100 hours. Then the obtained samples were ground and pressed into 13 mm diameter pellets under uniaxial stress of around 600 MPa using a hydraulic cold press for further studies. All the samples were handled in air after the heat treatment.

Structural characterization: X-ray diffraction (XRD) studies were performed on as synthesized materials and materials after the heat treatment in air. The powder XRD scans were carried out in a Bruker D8 Advance X-ray diffractometer with Cu K$_\alpha$ radiation in the Bragg-Brentano symmetric geometry. The sample stage was rotated at 15 rpm. The spectra was taken for a $2\theta$ range of 10° to 75° with a step of 0.015° and an integration time of 0.25 s.

Analytical characterization: Energy dispersive X-ray Spectroscopy (EDS) measurements were performed on as synthesized powders and powders after the heat treatment in air. The EDS spectra were obtained in a JEOL 7001F analytical field emission scanning electron microscope equipped with energy dispersive X-ray spectrometer. EDS spectra were taken on cold pressed pellets for the samples before heating, and on the pressed powder on carbon tape for samples after heating. All EDS spectra were acquired with settings of 15 kV accelerating voltage, around 67 μA emission current and a working distance of 15 mm. Different magnifications were used from 100× to 1000× to confirm the consistency across the powder collection. The spectra shown in Figure 5 and Figure 6 were recorded at 200×.

Raman spectroscopy: Raman spectroscopy measurements were performed in a Renishaw inVia confocal Raman Microscope with a 532 nm laser and a 20× objective lens. For samples before heating, Raman spectra were taken on the cold pressed pellets. To eliminate the possible signals from surface contaminations, measurements were performed right before and after sanding the surface with sand paper. The results were consistent. For samples after heating, the spectra were taken on loosely compacted powders.

Thermal stability test: Differential scanning calorimeter (DSC) and thermogravimetic analysis (TGA) measurements were performed simultaneously on a Netzsch STA 449 F3 Jupiter. Prior to each sample run, a correction was performed using both the reference and sample crucibles to account for any variations within the crucibles themselves. Sample powders obtained were weighed into alumina crucibles, equilibrated isothermally for 15 min, and heated to 1200 °C at a heating and cooling rate of 8 °C/min in air.

## III. Results and Discussion

### A. Structural diversity

Synthetic efforts of these ternary chalcogenides started more than half a century ago and their structural diversity has been studied experimentally and theoretically.[1,4-6,9,22-31] Due to the significantly larger size of chalcogen atoms than O atom, TMPCs display more structural distortions from the ideal cubic perovskite structure. Four major types of structural variations are of high interest, including distorted perovskite phase, needle-like phase, hexagonal perovskite phase, and the Ruddlesden-Popper (RP) phases. Schematics for these structural types are shown in Figure 1. The five compounds reported in this work cover two major structural types along with the RP phases.

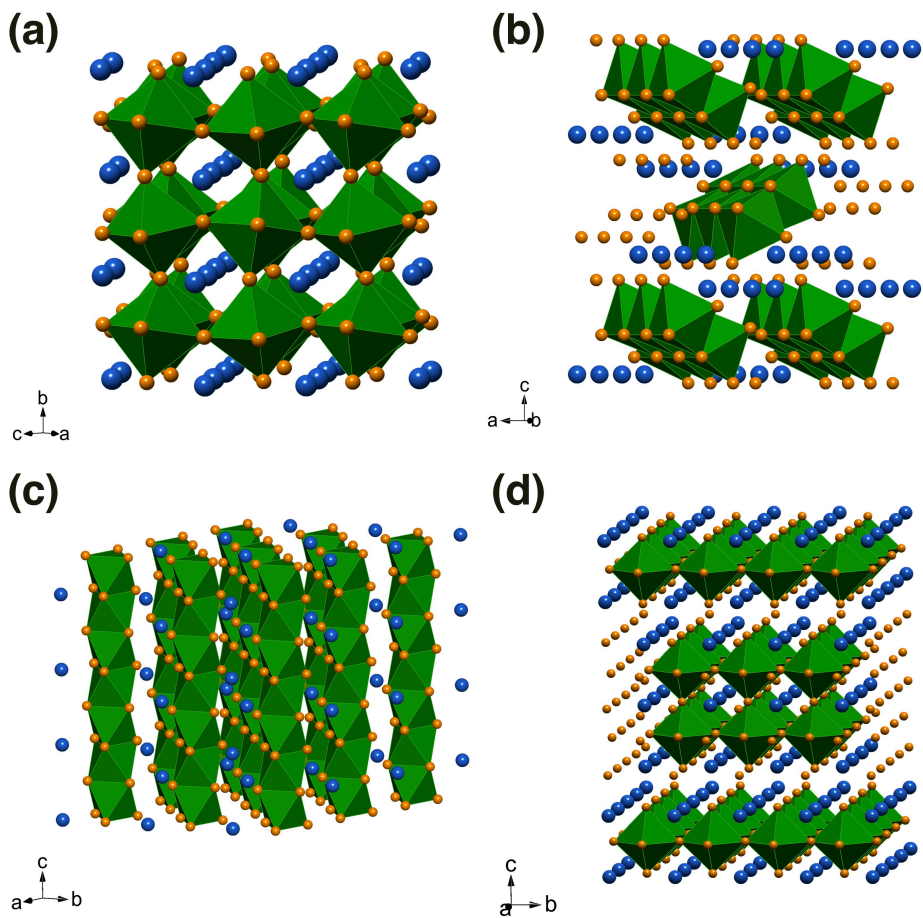

Figure 1. Schematics of various $ABX_3$ crystal structures for (a) distorted perovskite phase, (b) needle-like phase, (c) hexagonal perovskite phase, and (d) Ruddlesden-Popper phase ($n=2$). The blue and orange

spheres represent A site atoms and X site chalcogen atoms, respectively. The $BX_6$ octahedra are highlighted in green.

These structural variations can be viewed as different ways in which the $BX_6$ octahedra are connected in the extended three-dimensional (3D) space. The structure that is closest to the ideal cubic perovskite is the $GdFeO_3$ (GFO) structure with a typical space group $P$nma. The $BX_6$ octahedra are connected through a corner sharing 3D network (Figure 1(a)). Tilting of the octahedra breaks the cubic symmetry and ends up with a distorted perovskite phase with an orthorhombic structure. Among the compounds reported in this work, $BaZrS_3$ and higher temperature phase of $SrZrS_3$ ($\beta$-SZS) possess this structure. In the powder patterns of BZS and $\beta$-SZS (Figure 4(a), 4(b)), we can see most intense peaks qualitatively resemble peaks from the ideal perovskite structure, with weaker satellite peaks arising from the symmetry breaking due to the octahedra tilting. The lower temperature phase of $SrZrS_3$ ($\alpha$-SZS) is the needle-like phase with the $NH_4CdCl_3$ structure. In such a structure, the octahedra are connected by sharing edges, and two columns of edge-sharing octahedra form a chain. These parallel chains are extended along one direction and form a quasi-one-dimensional (Quasi-1D) structure (Figure 1(b)). The two structurally distinct phases can both be stabilized at room temperature by quenching from different growth temperatures.[9,30] Powder patterns of both SZS polymorphs are shown in Figure 4(b), which agree well with the previous structural study.[30] When smaller transition metals, such as Ti, sits in the B site, these compounds tend to stabilize in the $BaNiO_3$ structure, with a typical space group $P6_3/mmc$. $BX_6$ octahedra share opposing faces and form parallel single column chains. These chains are arranged in a hexagonal symmetry (Figure 1(c)). $BaTiS_3$ stabilizes in such a hexagonal perovskite phase. The Ti-Ti bond length is much shorted along the single column chains than across the chains. These strongly bonded $TiS_6$ chains are separated by Ba chains, forming a highly symmetric and yet also highly anisotropic Quasi-1D structure. Notably, the strong bonding along the chain also ensures that the natural cleavage of the crystal is along the chain direction, leading to a prismatic plane with easily accessible anisotropy. Such easily accessible in-plane anisotropy could lead to interesting anisotropic transport and physical properties. One example is the giant birefringence we have reported in $BaTiS_3$ very recently.[19] $Sr_{1+x}TiS_3$ (STS) has similar Quasi-1D chains with BTS, but the structure is more distorted and complex. STS samples adopt a composite crystal modulated structure, where the $TiS_6$ columns and Sr columns share common periodicities in the Basal plane, yet have different repeating patterns along the chain direction.[32-34] This leads to a slight off-stoichiometry of the composition, $Sr_{1+x}TiS_3$. The synthesis of STS and its anisotropic optical properties will be reported subsequently. RP phases of perovskites are formed by alternating a set number (*n*) of perovskite layers with the chemical formula $ABX_3$ and a rock salt layer AX. One schematic crystal structure for *n*=2 is shown in Figure 1(d). Such a 2D perovskite has a general formula of $A_{n+1}B_nX_{3n+1}$ for the case of the same cations in perovskite and rock salt layer. The RP phases could host interesting, different sets of octahedra tilting and distortion. Several theoretical studies explored the possibility of achieving static polarization in such materials to demonstrate bulk photovoltaic effect.[3,35] $Ba_3Zr_2S_7$ and $Ba_2ZrS_4$ are the *n*=2 and *n*=1 Ruddlesden-Popper phases of the perovskite sulfide $BaZrS_3$, respectively. The perovskite slabs with corner-sharing $ZrS_6$ octahedra are intercalated by one BaS layer, and are offset by half a unit cell along the face diagonal of the in-plane square lattice. Powder patterns of $Ba_3Zr_2S_7$ and $Ba_2ZrS_4$ are shown in Figure 4(a). We can see the signature low angle peaks corresponding to the larger lattice cell constants along the stacking directions.

## B. Thermal stability

Thermogravimetric analysis (TGA) and differential scanning calorimetry (DSC) were performed on the as synthesized samples in powder form to evaluate their thermal stability. All powders appear dark or dark brown colors before the heat treatment, as shown in Figure 2 (a). These powders are relatively stable in ambient conditions. There is no visible color change or measurable degradation over the course of one year. Around 30 mg of powders were used for each sample and gradually heated to 1200 °C in air. After the measurement, all samples turned into white powders, as shown in Figure 2(b).

**(a)**

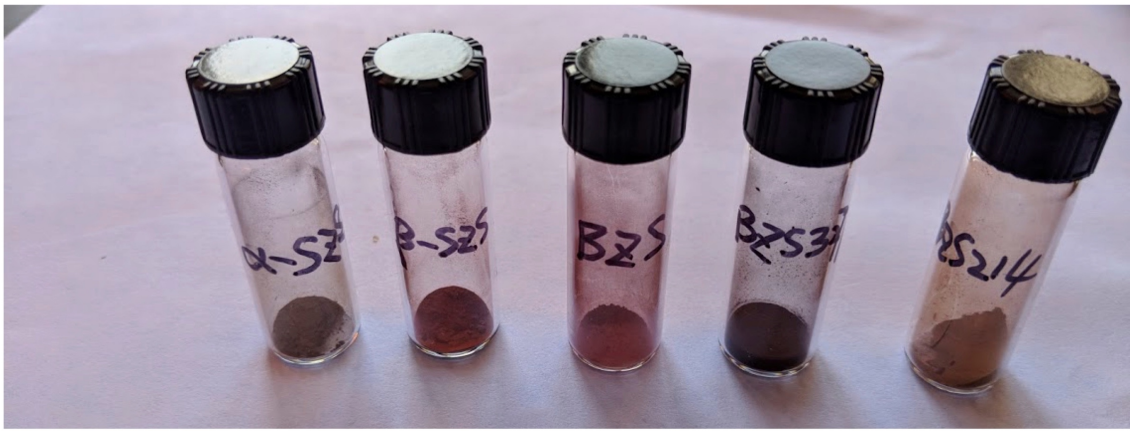

**(b)**

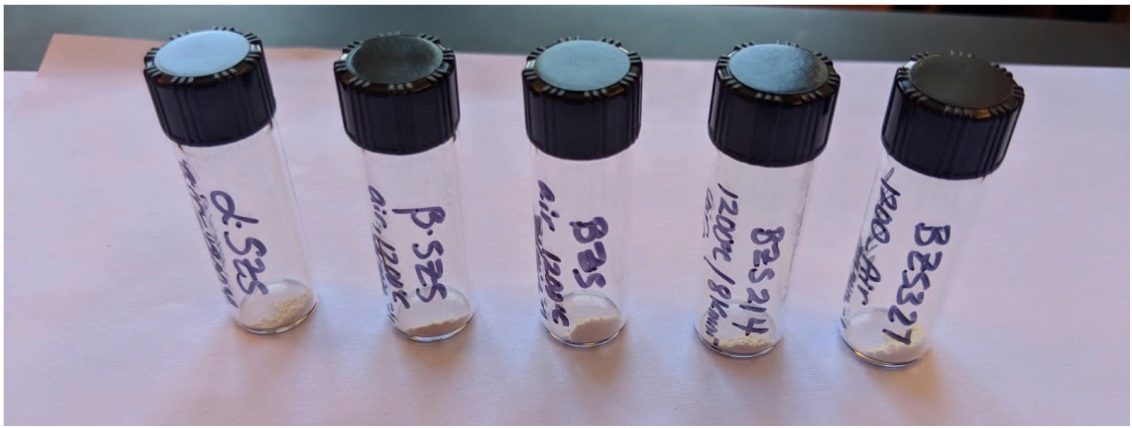

Figure 2. Optical pictures of five samples before (a) and after (b) heat treatment. The materials in (a) from left to right are α-SrZrS$_3$, β-SrZrS$_3$, BaZrS$_3$, Ba$_3$Zr$_2$S$_7$, and Ba$_2$ZrS$_4$. The materials in (a) from left to right are α-SrZrS$_3$, β-SrZrS$_3$, BaZrS$_3$, Ba$_2$ZrS$_4$, and Ba$_3$Zr$_2$S$_7$, respectively.

The weight change and DSC spectra as a function of the temperature are shown in Figure 3(a) and 3(b). We can see that a loss of weight in most samples occurs around 200 °C and there are corresponding small peaks in DSC spectra. We attribute this to the evaporation of iodine, which was used as catalysis in the sample synthesis. This was most obvious in α-SZS samples, as those samples needed slightly higher iodine concentrations to stabilize the desired needle-like phase. Apart from the loss of iodine in the powder mixture, all samples remain fairly stable in air until being heated well beyond 500 °C. The needle-like phase α-SZS is the first one subject to oxidize at ~550 °C with a sudden weight loss and a subsequent gradual weight gain. The oxidation of the two distorted orthorhombic phases, BZS and β-SZS happens at very close temperatures of just above 650 °C. And the two RP phases of BZS showed the highest stability with an oxidation onset a little below 800 °C. All these oxidations are signified by endothermic peaks in their corresponding DSC spectra. Another interesting feature in the TGA spectra is that α-SZS and BZS327 experienced relatively sharp weight loss followed by weight gain, while other materials experienced no appreciable weight loss and only gradual weight gain. The weight loss can be understood by the replacement of S with lighter O atoms. And the competing weight gain at higher temperatures is attributed to the formation of metal sulfates, as indicated by the XRD measurement of the powders after heating. The TGA and DSC data indicate that the lower symmetry needle-like phase α-SZS is the most vulnerable to degradation in elevated temperatures, and all the other higher symmetry phases, including distorted perovskite phases (BZS and $\beta$-SZS), and RP phases ($Ba_3Zr_2S_7$ and $Ba_2ZrS_4$) remain fairly stable in air up to well beyond 600 °C.

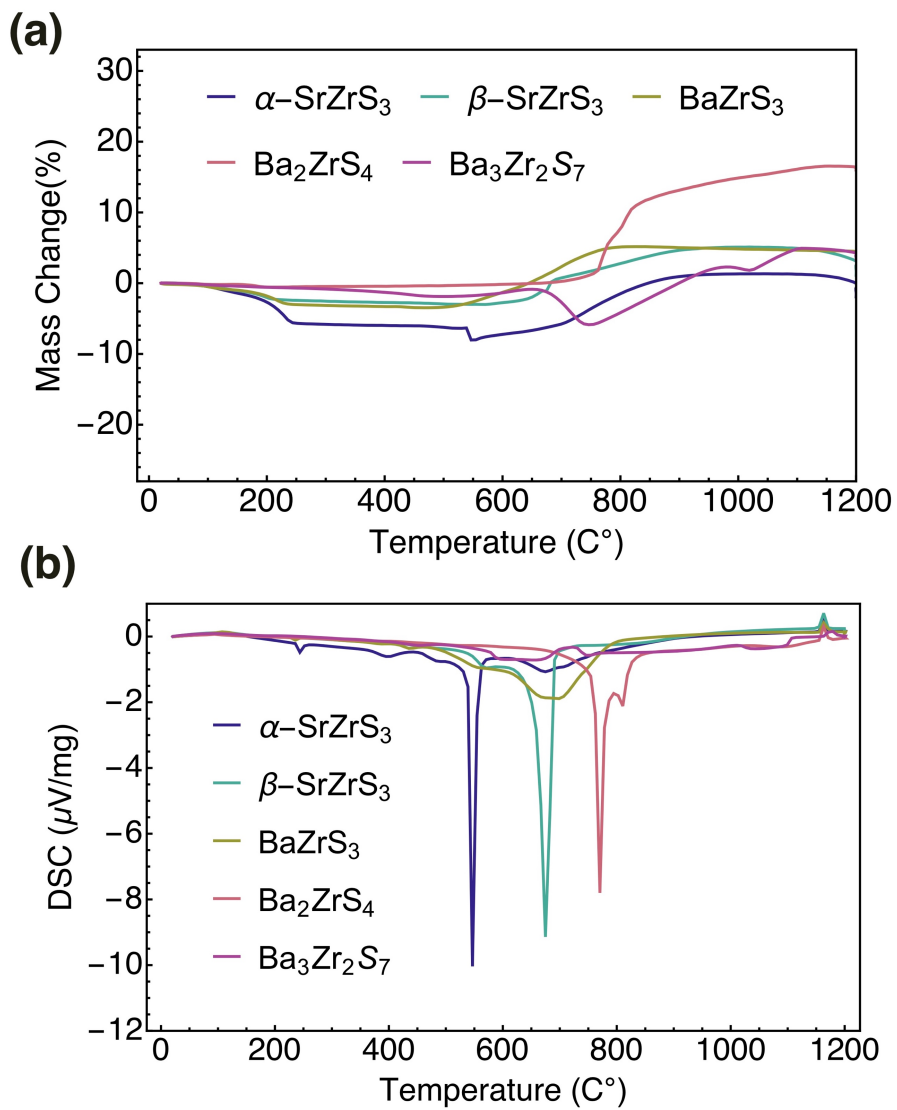

Figure 3. (a) Thermogravimetric analysis mass change and (b) differential scanning calorimetry of five samples as a function of the temperature.

## C. Oxidation products

To further evaluate the effect of heat treatment in air and understand the oxidation end products, we performed XRD, Raman spectroscopy and EDS studies on the samples before and after heat treatment. The comparison of the results is shown in Figure 4 - 9. As mentioned earlier, in the pre-treatment XRD, one can clearly see distinctive patterns for the SZS polymorphs and the BZS RP phases. However, the post-treatment XRD of two SZS polymorphs appear almost identical. The peaks can be attributed to a mixture of $SrZrO_3$, $SrSO_4$, and $ZrO_2$. After the heat treatment, BZS RP phases, with similar chemical compositions, showed very similar XRD patterns. The peaks can be attributed to a mixture of $BaSO_4$ and $ZrO_2$. Such oxidation results for these ternary chalcogenides are surprising and interesting, as it confirms the formation of sulfates for the A site cations and binary oxides for the B site cations, instead of the dominant formation of corresponding ternary oxides. The formation of sulfates also explained the unexpected weight gain at high temperatures.

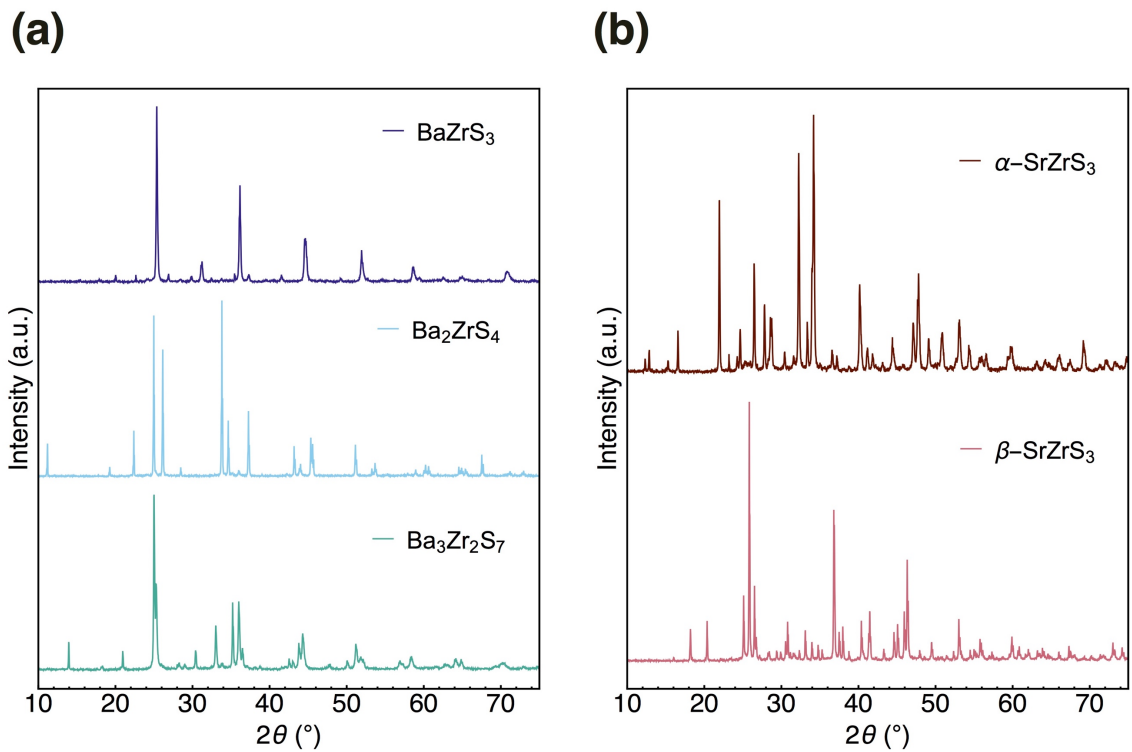

Figure 4. Powder XRD scans of all five samples before heat treatment.

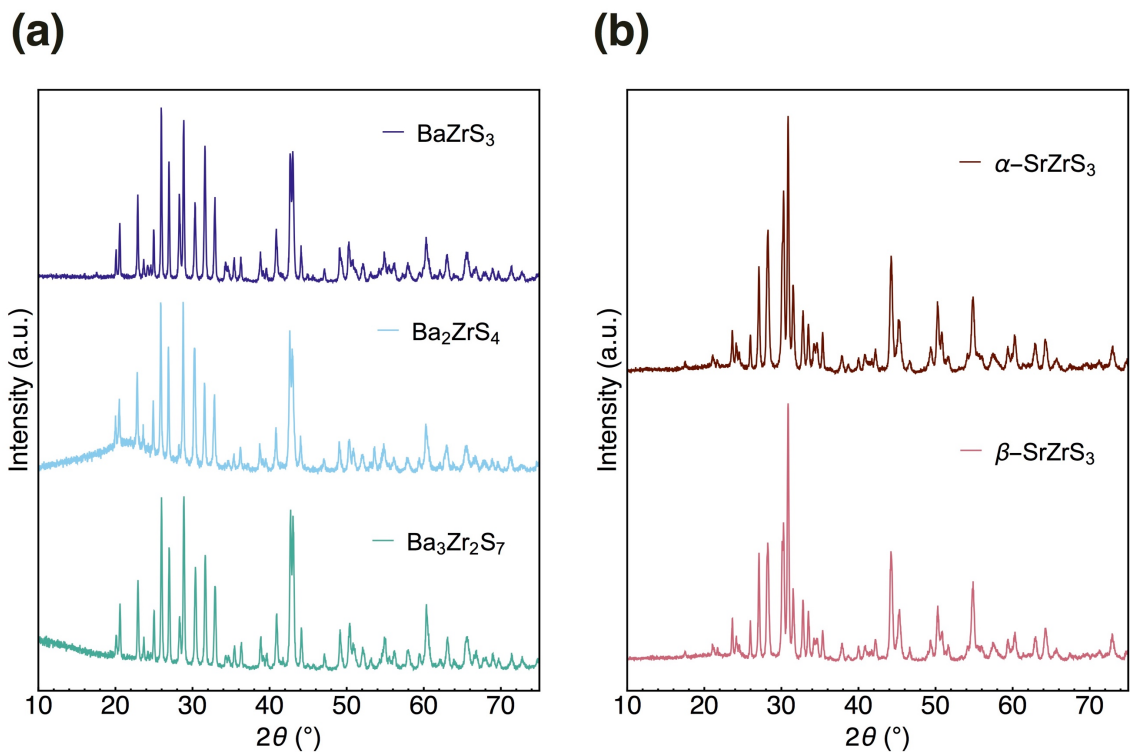

Figure 5. Powder XRD scans of all five samples after heat treatment.

Chemical composition study with EDS is in agreement with the structural study. Pre-treatment EDS spectra for all the materials showed expected composition with minimal O signal. Post-treatment EDS spectra showed much stronger O signal. Notably, S signal did not completely vanish, just significantly weaken, because of the sulfates formation.

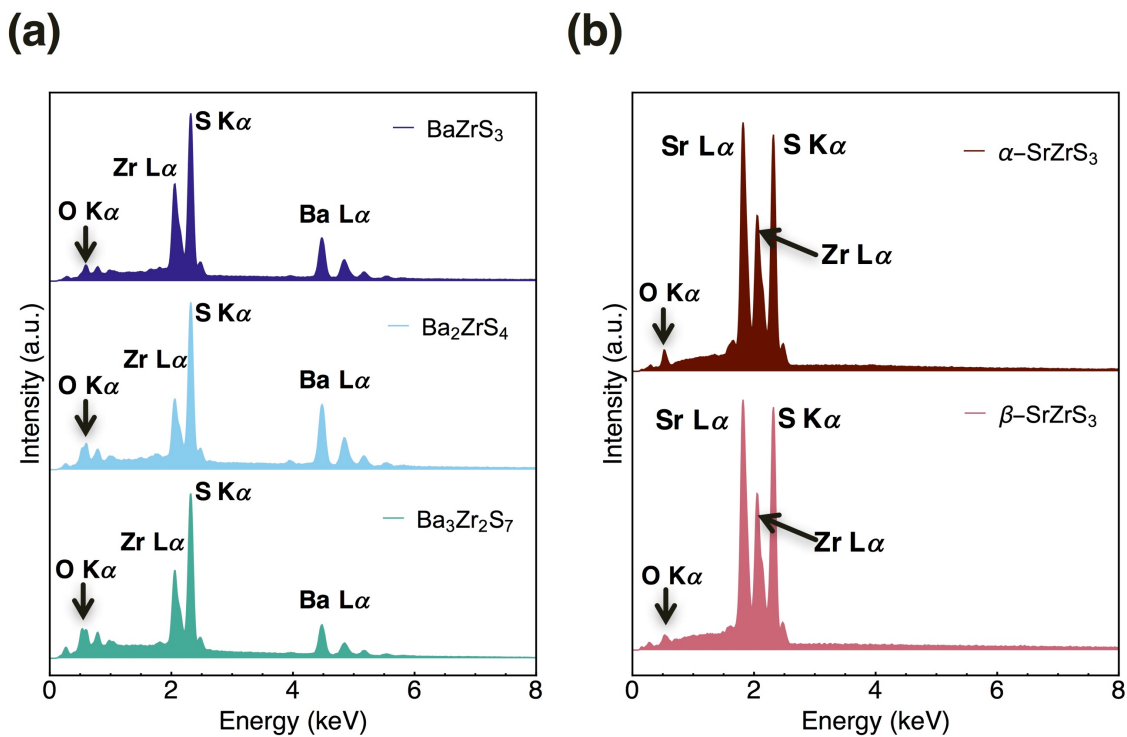

Figure 6. EDS spectra of all five samples before heat treatment.

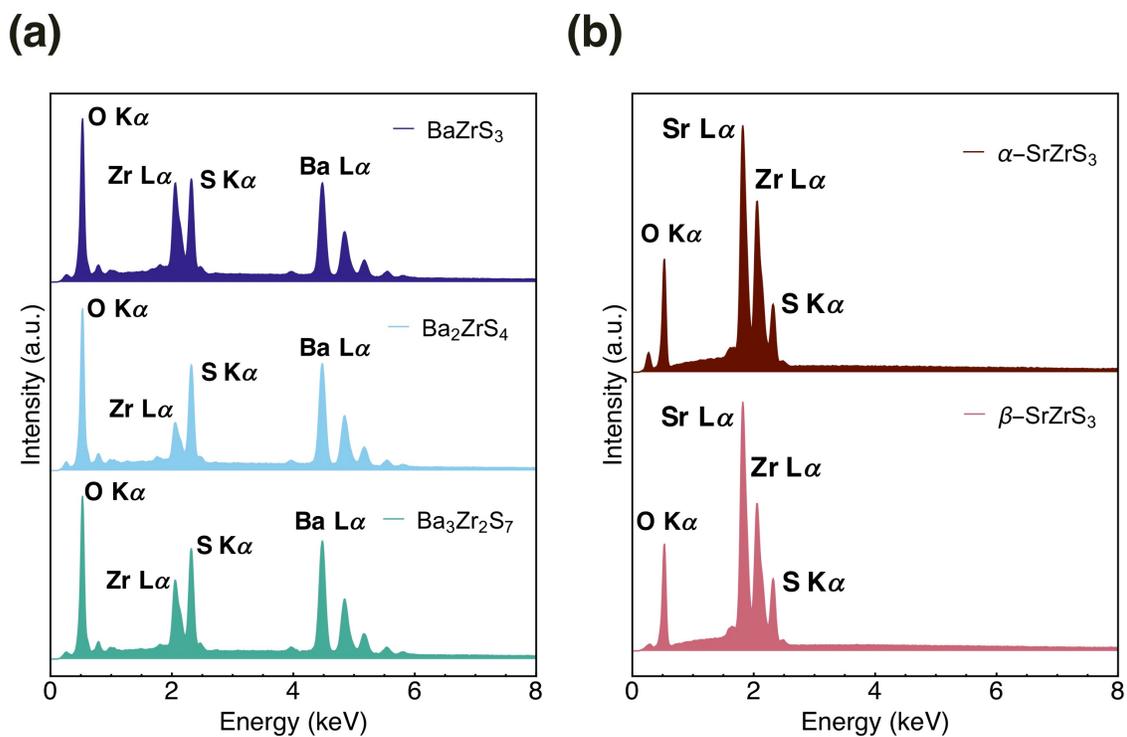

Figure 7. EDS spectra of all five samples after heat treatment.

The Raman spectra for any give material changed dramatically before and after heat treatment. The Raman spectra comparison across the materials revealed a similar trend with XRD. Pre-treatment samples showed distinctive Raman spectra due to different structures. BZS showed two signature $A_g$ peaks at ~130 cm$^{-1}$ and ~ 210 cm$^{-1}$, which agrees well with a previous extensive Raman study,[10] while Ba$_2$ZrS$_4$ and Ba$_3$Zr$_2$S$_7$ showed signature $A_{1g}$ Raman peak at ~305 cm$^{-1}$ and ~ 210 cm$^{-1}$, respectively. The post-treatment samples showed very similar Raman spectra, which agree well with the previously reported Raman spectra for the corresponding oxides and sulfates.[36-38] The spectra are dominated by peaks from monoclinic and tetragonal phases of ZrO$_2$, and the corresponding sulfates. Only two weak peaks at ~150 cm$^{-1}$ and ~ 410 cm$^{-1}$ in the post-treatment SZS polymorph Raman spectra can be attributed to the ternary oxide, SrZrO$_3$. The Raman spectra again confirmed the dominant oxidation products to be the B site metal binary oxides and A site metal sulfates.

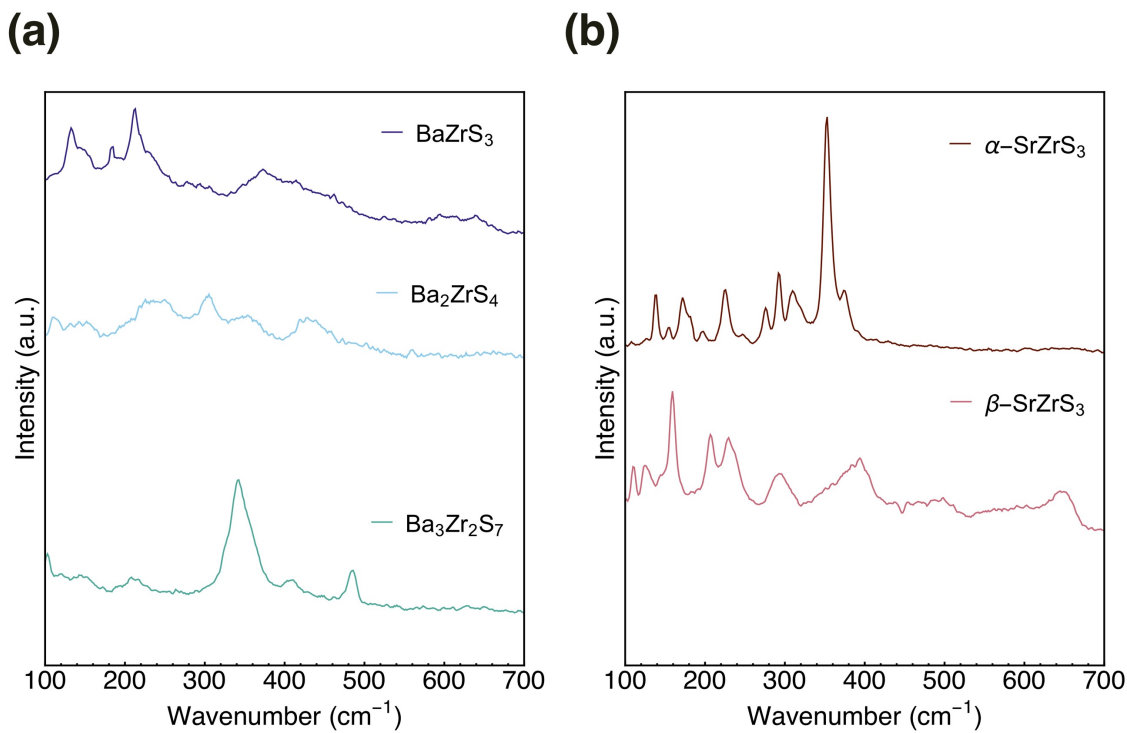

Figure 8. Raman spectra of all five samples before heat treatment.

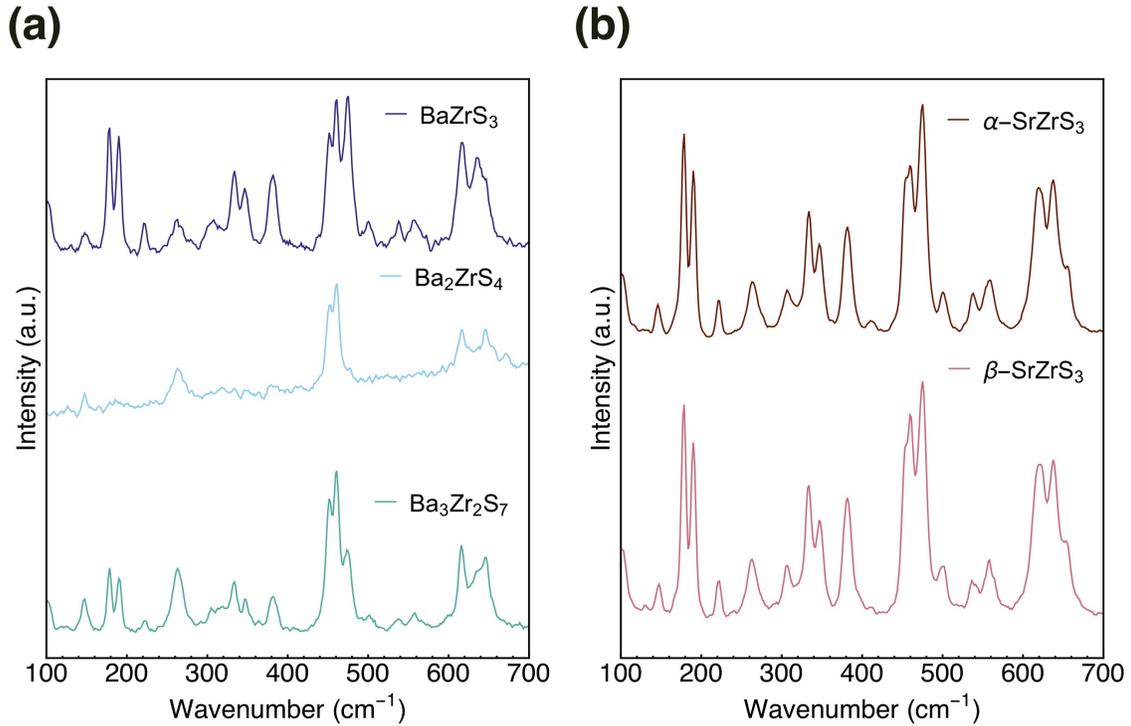

Figure 9. Raman spectra of all five samples after heat treatment.

## IV. Conclusion

This work has evaluated the thermal stability of several perovskite sulfides, including $BaZrS_3$ in distorted perovskite phase, two polymorphs of $SrZrS_3$ (needle-like phase and distorted perovskite phase), as well as two Ruddlesen-Popper phases, $Ba_2ZrS_4$ and $Ba_3Zr_2S_7$. All samples used in this work are synthesized in high quality polycrystalline form with solid state reaction in sealed quartz ampoules. Iodine was used to catalyze the reaction and reduce the synthesis time. TGA and DSC measurements were performed while these materials were heated to 1200 °C in air. The needle-like phase of $SrZrS_3$ is the one subject to degradation at the lowest temperature, 550 °C. All other phases appear pretty stable in air with an oxidation onset well beyond 600 °C. XRD, EDS and Raman studies confirmed the dominant oxidation products as a mixture of A site metal sulfates (SrSO$_4$, BaSO$_4$) and B site metal oxides, ZrO$_2$. The desirable high thermal stability of this class of materials opens up possibilities for alternative thermoelectric materials with earth benign and abundant composition. Furthermore, we believe that the rich tunability from structural diversity and vast chemical composition in these transition metal perovskite chalcogenides offers an exciting platform to realize innovative, desired functionalities for energy and optoelectronic applications in general.

## Acknowledgments

This work is supported by USC Viterbi School of Engineering Startup Funds and the Air Force Office of Scientific Research under award number FA9550-16-1-0335. S.N. acknowledges Link Foundation Energy Fellowship. J.M.G. and B.C.M. gratefully acknowledge support from the Office of Naval Research Grant No. N00014-15-1-2411.

## References

1. Sun, Y.-Y., Agiorgousis, M. L., Zhang, P. & Zhang, S. Chalcogenide Perovskites for Photovoltaics. *Nano Lett.* **15,** 581–585 (2015).
2. rbel, S. K. X., Marques, M. A. L. & Botti, S. Stability and electronic properties of new inorganic perovskites from high-throughput ab initio calculations. *J. Mater. Chem. C* **4,** 3157–3167 (2016).
3. Wang, H., Gou, G. & Li, J. Ruddlesden–Popper perovskite sulfides A$_3$B$_2$S$_7$: A new family of ferroelectric photovoltaic materials for the visible spectrum. *Nano Energy* **22,** 507–513 (2016).
4. Ju, M.-G., Dai, J., Ma, L. & Zeng, X. C. Perovskite chalcogenides with optimal bandgap and desired optical absorption for photovoltaic devices. *Adv. Energy Mater.* **48,** 1700216 (2017).
5. Nijamudheen, A. & Akimov, A. V. Criticality of Symmetry in Rational Design of Chalcogenide Perovskites. *J. Phys. Chem. Lett.* **9,** 248–257 (2017).
6. Kuhar, K. *et al.* Sulfide perovskites for solar energy conversion applications: computational screening and synthesis of the selected compound LaYS$_3$. *Energy Environ.*

*Sci.* **10,** 2579–2593 (2017).
7. Meng, W. *et al.* Alloying and Defect Control within Chalcogenide Perovskites for Optimized Photovoltaic Application. *Chem. Mater.* **28,** 821–829 (2016).
8. Perera, S. *et al.* Chalcogenide perovskites – an emerging class of ionic semiconductors. *Nano Energy* **22,** 129–135 (2016).
9. Niu, S. *et al.* Bandgap Control via Structural and Chemical Tuning of Transition Metal Perovskite Chalcogenides. *Adv. Mater.* **29,** 1604733 (2017).
10. Gross, N. *et al.* Stability and Band-Gap Tuning of the Chalcogenide Perovskite $BaZrS_3$ in Raman and Optical Investigations at High Pressures. *Phys. Rev. Appl* **8,** 044014 (2017).
11. Wang, J. & Kovnir, K. Giant anisotropy detected. *Nat. Photon.* **508,** 373 (2018).
12. He, J., Liu, Y. & Funahashi, R. Oxide thermoelectrics: The challenges, progress, and outlook. *Journal of Materials Research* **26,** 1762–1772 (2011).
13. Muta, H., Kurosaki, K. & Yamanaka, S. Thermoelectric properties of rare earth doped SrTiO3. *J. Alloy Compd.* **350,** 292–295 (2003).
14. Ohta, H., Sugiura, K. & Koumoto, K. Recent progress in oxide thermoelectric materials: p-type $Ca_3CO_4O_9$ and n-type $SrTiO_3$. *Inorg. Chem.* **47,** 8429–8436 (2008).
15. WEBER, W. J., GRIFFIN, C. W. & BATES, J. L. Effects of Cation Substitution on Electrical and Thermal Transport Properties of $YCrO_3$ and $LaCrO_3$. *J. Am. Ceram. Soc.* **70,** 265–270 (1987).
16. Ohtaki, M., Koga, H., Tokunaga, T., Eguchi, K. & Arai, H. Electrical Transport Properties and High-Temperature Thermoelectric Performance of $(Ca_{0.9}M_{0.1})MnO_3$ (M = Y, La, Ce, Sm, In, Sn, Sb, Pb, Bi). *J. Solid State Chem.* **120,** 105–111 (1995).
17. Okuda, T., Nakanishi, K., Miyasaka, S. & Tokura, Y. Large thermoelectric response of metallic perovskites: Sr1-xLaxTiO3  (0~x~0.1). *Phys. Rev. B* **63,** 113104 (2001).
18. Yasukawa, M. & Murayama, N. A promising oxide material for high-temperature thermoelectric energy conversion: $Ba_{1-x}Sr_xPbO_3$ solid solution system. *Mater Sci Eng B Solid State Mater Adv Technol* **54,** 64–69 (1998).
19. Niu, S. *et al.* Giant optical anisotropy in a quasi-one-dimensional crystal. *Nat. Photon.* **12,** 392–396 (2018).
20. Niu, S. *et al.* Mid-wave and Long-wave IR Linear Dichroism in a Hexagonal Perovskite Chalcogenide. *arXiv* eprint arXiv:1806.09688, (2018).
21. Niu, S. *et al.* Ideal Bandgap in a 2D Ruddlesden-Popper Perovskite Chalcogenide for Single-junction Solar Cells. *arXiv* eprint arXiv:1806.09685, (2018).
22. Brehm, J. A., Bennett, J. W., Schoenberg, M. R., Grinberg, I. & Rappe, A. M. The structural diversity of $ABS_3$ compounds with $d^0$ electronic configuration for the B-cation. *J. Chem. Phys.* **140,** 224703 (2014).
23. Lelieveld, R. & Ijdo, D. J. W. Sulphides with the GdFeO3 structure. *Acta Cryst. B* **36,** 2223–2226 (1980).
24. Clearfield, A. The synthesis and crystal structures of some alkaline earth titanium and zirconium sulfides. *Acta Cryst.* **16,** 135–142 (1963).
25. Hahn, H. & Mutschke, U. Untersuchungen über ternäre Chalkogenide. XI. Versuche zur Darstellung von Thioperowskiten. *Z. Anorg. Allg. Chem.* **288,** 269–278 (1957).
26. Huster, J. Die Kristallstruktur von $BaTiS_3$. *Z. Naturforsch. B* **35,** (1980).
27. Bin Okai, Takahashi, K., Saeki, M. & Yoshimoto, J. Preparation and crystal structures of some complex sulphides at high pressures. *MRS Bull.* **23,** 1575–1584 (1988).
28. Tranchitella, L. J., Chen, B. H., Fettinger, J. C. & Eichhorn, B. W. Structural evolutions in

29. Aleksandrov, K. S. & BartolomÉ, J. Structural distortions in families of perovskite-like crystals. *Phase Transitions* **74,** 255–335 (2001).
30. Lee, C.-S., Kleinke, K. M. & Kleinke, H. Synthesis, structure, and electronic and physical properties of the two SrZrS3 modifications. *Solid State Sci.* **7,** 1049–1054 (2005).
31. Bennett, J. W., Grinberg, I. & Rappe, A. M. Effect of substituting of S for O: the sulfide perovskite BaZrS$_3$ investigated with density functional theory. *Phys. Rev. B* **79,** 235115 (2009).
32. Tranchitella, L. J., Fettinger, J. C., Dorhout, P. K., Van Calcar, P. M. & Eichhorn, B. W. Commensurate columnar composite compounds: synthesis and structure of Ba$_{15}$Zr$_{14}$Se$_{42}$ and Sr$_{21}$Ti$_{19}$Se$_{57}$. *J. Am. Chem. Soc.* **120,** 7639–7640 (1998).
33. Gourdon, O., Petricek, V. & Evain, M. A new structure type in the hexagonal perovskite family; structure determination of the modulated misfit compound Sr$_{9/8}$TiS$_3$. *Acta Cryst.* **B56,** 409–418 (2000).
34. Gourdon, O. *et al.* Influence of the Metal–Metal Sigma Bonding on the Structures and Physical Properties of the Hexagonal Perovskite-Type Sulfides Sr$_{9/8}$TiS$_3$, Sr$_{8/7}$TiS$_3$, and Sr$_{8/7}$[Ti$_{6/7}$Fe$_{1/7}$]S$_3$. *J. Solid State Chem.* **162,** 103–112 (2001).
35. Zhang, Y., Shimada, T., Kitamura, T. & Wang, J. Ferroelectricity in Ruddlesden–Popper Chalcogenide Perovskites for Photovoltaic Application: The Role of Tolerance Factor. *J. Phys. Chem. Lett.* **8,** 5834–5839 (2017).
36. Quintard, P. E., Barberis, P., Mirgorodsky, A. P. & Merle-Mejean, T. Comparative lattice-dynamical study of the Raman spectra of monoclinic and tetragonal phases of zirconia and hafnia. *J. Am. Ceram. Soc.* **85,** 1745–1749 (2002).
37. DAWSON, P., HARGREAVE, M. M. & WILKINSON, G. R. Polarized Ir Reflection, Absorption and Laser Raman Studies on a Single-Crystal of Baso4. *Spectrochimica Acta. A* **33,** 83–93 (1977).
38. Kamishima, O., Hattori, T., Ohta, K., Chiba, Y. & Ishigame, M. Raman scattering of single-crystal SrZrO3. *J. Phys. Condens. Matter* **11,** 5355–5365 (1999).

the Sr$_{1-x}$Ba$_x$ZrSe$_3$ series. *J. Solid State Chem.* **130,** 20–27 (1997).